\documentclass[usenatbib]{mn2e}
\pdfoutput=1

\usepackage[pdftex]{graphicx}

\begin{document}

\title[Dark matter merging induced turbulence]{Dark matter merging induced turbulence as an efficient engine for gas cooling}

\author[Prieto, Jimenez \& Mart\'i]{Joaquin Prieto$^1$\thanks{email:joaquin.prieto.brito@gmail.com}, Raul Jimenez$^{2,1}$, Jose Mart\'i$^1$\\
$^1$ICC, Universitat de Barcelona (IEEC-UB), Marti i Franques 1, E08028, Barcelona, Spain\\
$^2$ICREA}

\maketitle

\begin{abstract}
We have performed a cosmological numerical simulation of primordial baryonic gas 
collapsing onto a $3\times10^7$M$_{\odot}$ dark matter (DM) halo. We show that the 
large scale baryonic accretion process and the merger of few $\sim10^6$ M$_{\odot}$ DM 
halos, triggered by the gravitational potential of the biggest halo, is enough to create 
super sonic ($\mathcal{M}>10$) shocks and develop a turbulent environment. In this 
scenario the post shocked regions are able to produced both H$_2$ and HD molecules 
very efficiently reaching maximum abundances of $n_\mathrm{H_2}\sim10^{-2}n_\mathrm{H}$ 
and $n_\mathrm{HD}\sim \mathrm{few}\times10^{-6}n_\mathrm{H}$, enough to cool the gas below 
100K in some regions. The kinetic energy spectrum of the turbulent primordial gas is close 
to a Burgers spectrum, $\widehat{E}_k\propto k^{-2}$, which could favor the formation of low mass primordial stars. The solenoidal to total kinetic 
energy ratio is $0.65\la R_k\la0.7$ for a wide range of wave numbers; this value is close 
to $R_k\approx 2/3$ natural equipartition energy value of a random turbulent flow. 
In this way turbulence and molecular cooling seem to work together in order to produce 
potential star formation regions of cold and dense gas in primordial environments. We conclude 
that both the mergers and the collapse process onto the main DM halo provide enough energy 
to develop super sonic turbulence which favor the molecular coolants formation: this 
mechanism, which could be universal and the main route toward formation of the first 
galaxies, is able to create potential star forming regions at high redshift.
\end{abstract}


\section{Introduction}
In the currently accepted paradigm to describe the universe, LCDM, the formation of bound 
dark matter structures takes place via hierarchical growth. While this build up is driven 
by dark matter accretion onto an already virialized dark matter halo, the process of 
merging is equally important. The effect of dark matter mergers in building up the dark 
matter halo has been studied in enormous detail both numerically and analytically. Due to 
new hydrodynamical simulations that have achieved enough resolution, the effect of merging 
of baryons is being taken into account, including also the important process of transfer of baryonic matter between dark matter merging halos. 

One aspect of merging that has not been studied in detail is the development of (supersonic) 
turbulence in the merging halos due entirely to the dark matter merging process and its 
consequent effect on the environment. The current paradigm to power turbulence in the 
interstellar medium is mainly through the process of SN explosions \citep{NormanFerrara1996}. 
However, it is possible that the dark matter merging process itself does power the initial phase 
of turbulence and thus star formation. 

Using numerical simulations from cosmological initial condition \citet{WiseAbel2007} and 
\citet{Greifetal2008} have discussed the generation of turbulent motions in primordial gas 
through the virialization process of halos of mass $\approx 10^8$M$_{\odot}$ and 
$\approx 5\times 10^7$M$_{\odot}$, respectively. They have shown that 
%the accretion of 
%minihalos and the collapse of primordial gas on such halos, particularly the cold accretion 
%through filaments, generates 
supersonic turbulent motions, which partially ionize the 
primordial gas allow an efficient formation of H$_2$ and HD molecules and efficient 
cooling to a gas temperature below $\sim100$K in some regions. They have argued that these 
turbulent low temperature regions could be sites of efficient star formation. 

The result of \citet{Greifetal2008} is somehow different from the conclusions of \cite{JohnsonBromm2006} 
who have analytically shown that in order to ionize the primordial gas through the DM virialization 
process a halo mass above $10^8$M$_{\odot}$ is needed. This apparent contradiction is easily 
understood because \cite{JohnsonBromm2006} took into account the fixed halo average circular 
velocity for the velocity shock waves instead a velocity distribution for primordial gas (which 
can take values higher and lower than the average circular velocity). Summarizing, \cite{Greifetal2008} 
have shown that the DM virialization process of halos with mass $\ga10^7$M$_{\odot}$ can generate a 
supersonic turbulent media (triggered by both hot and cold gas accretion) where coolants (H$_2$ and HD molecules) are generated efficiently. 
%However, the two numerical experiment mentioned above have some limitations in resolution in 
%order to investigate DM halos with smaller masses, which are the dominant population at $z > 10$. They 
%also have not the numerical resolution needed to fully characterize the turbulence.
% Second main point
However, despite of the interesting insights on the development of turbulence in cosmological simulations of primordial gas made by the simulation mentioned above, their authors have not pay attention on the impact of the halo merger process as a turbulence inductor and they have related the onset of turbulence just with the gas accretion onto the DM halos. 

On the other hand, \citet{WiseAbel2007} mentioned that the minor and major mergers are able to produce turbulence by Kelvin-Helmholtz instabilities, as well as by gas accretion. Despite of this mention, the both studies mentioned above neither characterize the turbulence with power spectrum in protogalaxy environments nor characterize the gas conditions with probability distribution functions in order to have a clear idea about the primordial gas conditions inside the first protogalaxies. No previous study has ever done this in such kind of environments, therefore we believe our study brings many new results to the field.

In the last years there have been a number of works studying the development of turbulence under
isothermal and non isothermal conditions. A very interesting aspect of supersonic gas turbulence
is the effect of the gas dynamic on the gas chemistry (mainly on the formation of gas coolants).
\citet{Milo2011} have shown that compressive forcing turbulence produces  H$_2$ formation 
faster than in the solenoidal forcing case. This behavior is explained by these authors due to 
the broader compressive forcing density PDF which creates more H$_2$ molecules at initial times. 
The results of \citet{Milo2011} could be placed in a cosmological context and argue that if a DM 
halo is able to produce enough compressive supersonic turbulent motions it is possible to produce a 
large amount of molecules and cool down the primordial gas very efficiently. 

In this paper we investigate in detail how turbulence is generated by both mini DM halo merging and gas 
collapse using a cosmological size simulation. We characterize the gas physical conditions computing probability distribution functions of gas (density, temperature and velocity) and the gas velocity power spectrum of this cosmological simulation, which is an study not done and can be useful for idealized primordial star formation simulations.
We follow the merging process in detail and investigate the onset of a turbulent medium 
inside the merging halos, including very small mass halos ($M \sim 10^6M_{\odot}$). The most interesting findings of our work are that merging does indeed power 
turbulence, but most interesting is that it enhances significantly the coolants in the primordial 
gas, thus significantly lowering the temperature of the primordial gas, even below $100$ K in 
some regions. Furthermore, we found that the velocity power spectrum is nearby a Burgers spectrum $\propto k^{-2}$ and the solenoidal to total kinetic energy ratio is nearby 2/3 as found in idealized random fluids simulations.
%We emphasize that our gas characterization can be used as initial condition of idealized hydrodynamical simulation of primordial star formation.  

The merger inducing turbulence process can be universal and can also take place at lower redshift, thus powering the initial phases of star formation in merging halos. The paper is structured as follows: in \S~2 we describe the methodology and follow in \S~3 with our main results and analysis. We continue with a discussion of our results in \S~4 and conclude and summarize in \S~5.

\section{Methodology}
We used the adaptive mesh refinement (AMR) code RAMSES \citep{Teyssier2002} with a modified 
non-equilibrium cooling module with 21 species (including H$_2$, HD, and LiH molecules
with their cooling functions) in order to follow the chemo-thermal evolution of primordial 
gas and to study the gas gravitational collapse process in pristine environments \citep{jpp}. All chemical
reaction rates were taken from \citet{Stanciletal1996}, \citet{GP98}, and \citet{GA08}.

We worked in a concordance $\Lambda$CDM cosmological model: $h=0.72$, $\Omega_{\Lambda}=0.73$,
$\Omega_{m}=0.27$, $\Omega_{b}=0.04$, $\sigma_8=0.9$ and $n_s=0.95$. The dynamical initial
conditions were taken from mpgrafic \citep{Prunetetal2008} and the initial chemical abundances
were taken from \citet{GP98} at $z=120$ (the initial redshift of the simulation).

We first simulated a volume of (1~Mpc/h)$^3$ of only DM, with 256$^3$ particles. Using the HOP
algorithm \citep{EisensteinHut1998}, we identified a $3\times 10^7$~M$_{\odot}$ halo at $z=10$.
In order to better resolve the formation of that halo and to study its baryon component, we 
re-simulated the same (1Mpc/h)$^3$ volume with 512$^3$ particles (with a particle mass of 
approximately $\sim$800M$_{\odot}$), including gas with non-equilibrium primordial chemistry
from the beginning of the simulation at z=120. The gas dynamics was computed on a root grid of 
64$^3$ computational zones, which was geometrically refined towards the center, increasing the
spatial resolution by a factor of two within the central 1/8 of the volume. This refinement
criterion was applied three times, generating 4 nested meshes (including the root grid), each
with 64$^3$ computational cells. 

Inside the innermost 64$^3$ mesh, the gas dynamics was computed using seven extra levels of
adaptive mesh refinement by a factor of two. This adaptive refinement was based on four 
different criteria: i) Lagrangian refinement based on the number density of DM particles (a 
finer refinement level is created in cells containing more than 4 DM particles), ii) Lagrangian
refinement based on the baryonic mass density,  iii) refinement based on the gas pressure
gradient (for $\Delta p/p\ge2$), and iv) refinement based on the Jeans' length, to satisfy 
Truelove's condition \citep{Trueloveetal1997}. The pressure gradient criterion was included 
in order to better resolve the turbulent flow, as discussed in \cite{Kritsuketal2006}. The
three geometrical refinement levels, plus the additional seven adaptive refinement levels, give 
an effective spatial resolution corresponding to that achieved by a uniform mesh of $65,536^3$
computational elements, and corresponding to a proper size of $\approx$1.95pc at $z\approx10.0$.

This setup was used from the beginning of the simulation, at $z=120$, until $z=50$. At $z=50$,
another geometrical refinement criterion was added (while maintaining the others): Uniform
spatial resolution till level 15 was imposed around the densest central region of the halo, 
creating a uniform mesh of 512$^3$ elements which was refined with a $J=0.125$ Jeans criteria
till level 16 (due to technical reasons the uniform refinement criteria was imposed gradually 
and it was completed at z=30). This high refinement region, covering a proper size of 
$\approx$2.0 kpc at $z\approx10.0$, served the purpose of better resolving the generation 
of turbulent motions in the central region of the halo. The results presented in this work 
are based on the analysis of this central 2.0kpc region at the end of the simulation at 
$z\approx10.0$. 

\section{Results and Analysis}
The simulation presented in this work follows the baryonic matter accretion process onto 
a $\approx3\times10^7$M$_\odot$ (hereafter the main halo) with $T_{\mathrm{vir}}\approx7770$ 
K and $V_{\mathrm{circ}}\approx14.6$ km/s. 

At $z\approx14$ the main halo is crossed by two interacting DM minihalos of $\sim10^6$M$_\odot$. These two interacting minihalos merge and create a spinning baryonic over-density going 
out the main halo central region. Due to this violent merging process the gas develops a shock 
wave which increases the free electron fraction. The post shocked gas develops turbulence and 
creates H$_2$ and HD molecules very efficiently behind the merged over-density. 
 
After the merging process the main halo continues accreting gas mainly by filamentary flows.
The accretion process heats up the gas till $T\ga 7\times10^3$K and accelerates it developing 
super sonic shocks. Near $z\approx12$, at the densest main halo central regions the post 
shocked primordial gas has created a lot of both H$_2$ and HD molecules in a very efficient 
way. Furthermore, due to the compressional turbulent motions triggered by the roughly radial 
collapse, the molecular formation is enhanced as well. Because of the molecular formation process 
the turbulent gas can reach temperatures even below 100 K in some regions and it is able to 
form potential star formation regions of dense gas ($n\ga10^4$cm$^{-3}$) at low temperatures 
($T\sim \mathrm{few}\times100$K).

\begin{figure*}
\includegraphics[height=14cm,width=15cm]{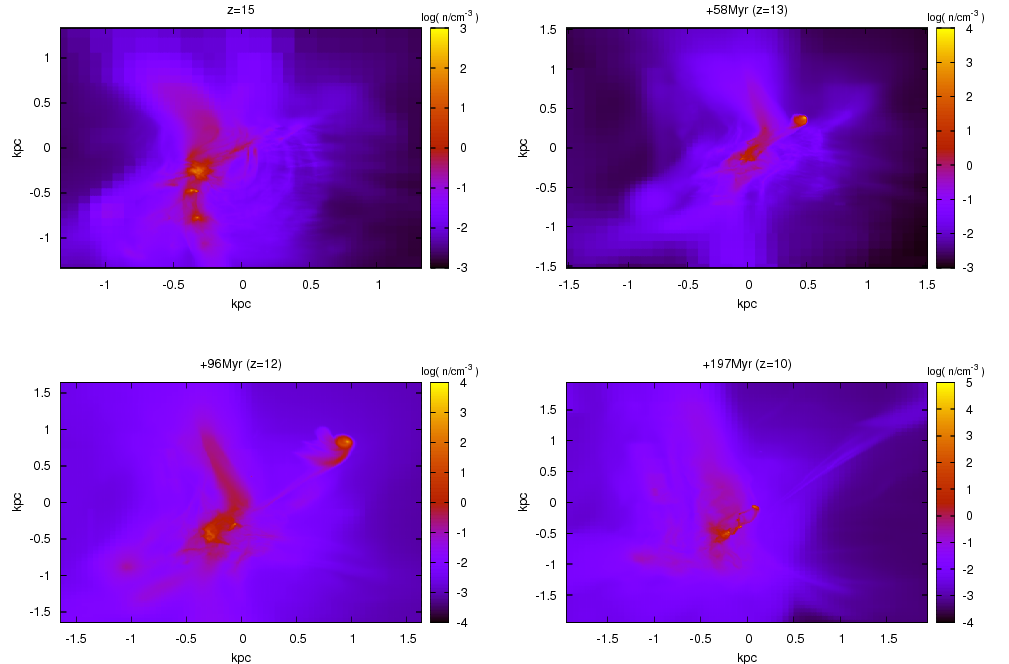}
\caption{The z projection of the mass weighted gas number density at z=15 (top left), 13 (top right), 12 (bottom left) and 10 (bottom right). Top left: At $z=15$, near the main halo central region, appear three interacting mini halos; the nearest two over-densities will merge. Top right: At $z=13$, after the merger a spinning over-density crosses the main halo central region as a super sonic bullet; the over-density leaves a post shock region where molecular coolants are formed very efficiently due to both the turbulent motions and the enhancement in the free electron fraction. Bottom left: At $z=12$ the spinning over-density is leaving the main halo central region which starts to develop more over-densities due to the gas accretion process. Bottom right: At $z=10$ the gas has developed a number of over-densities near the main halo central region as a consequence of the combined effect of molecular cooling and gas dynamics in turbulent compressed regions.}
\label{evolution1}
\end{figure*}

\begin{figure*}
\centering
\includegraphics[height=14cm,width=15cm]{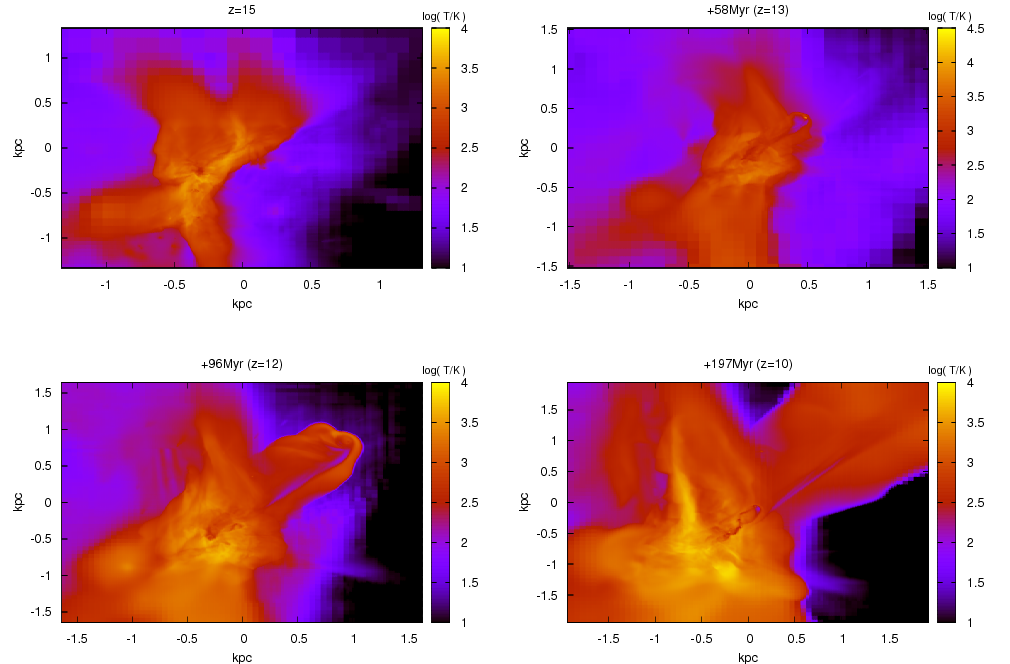}
\caption{The z projection of the mass weighted gas temperature. The different maps correspond to the same redshifts of figure \ref{evolution1}. The time evolution shows how after the merger a cold tail is formed behind the spinning over-density. This remaining turbulent cold gas is fed by the accretion near the main halo central region and is able to develop dense and cold over-densities.}
\label{evolution2}
\end{figure*}

\begin{figure*}
\centering
\includegraphics[height=14cm,width=15cm]{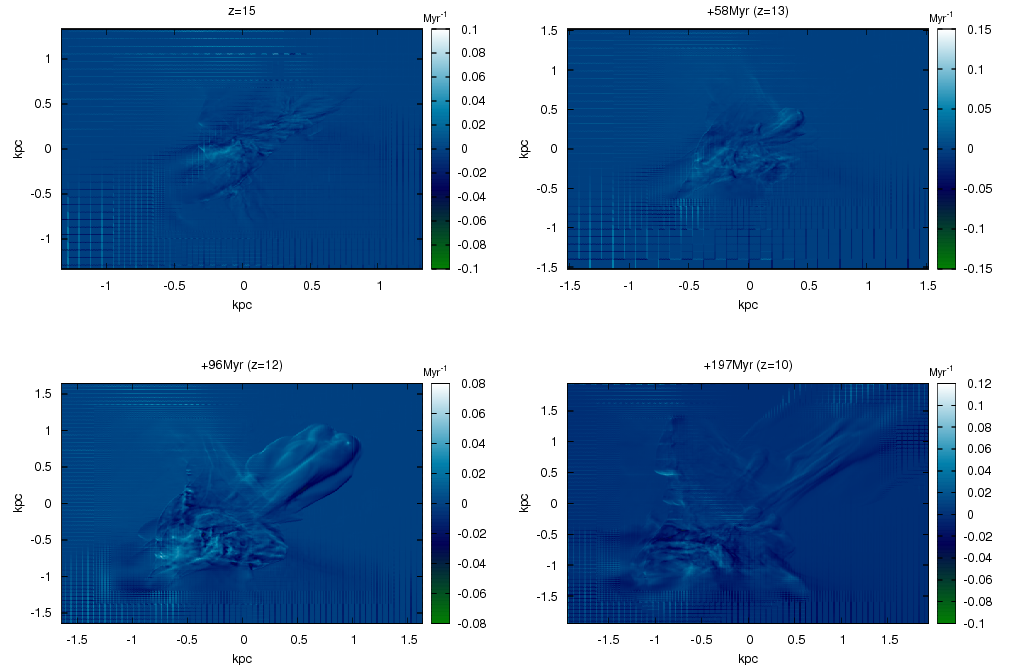}
\caption{The z projection of the z component of gas vorticity $(\nabla\times\vec{v})_z$. The different maps correspond to the same redshifts of figure \ref{evolution1}. The maps show how a turbulent environment is developed by both the merger an the accretion process near the main halo central region.}
\label{evolution3}
\end{figure*}

Figures \ref{evolution1}, \ref{evolution2} and \ref{evolution3} show the z projection of the gas 
number density, gas temperature and gas vorticity inside a cubic box of $\sim$3 kpc of side
(larger than the highest refinement level box of $\sim2$ kpc) 
as a function of redshift. The horizontal and vertical lines in the vorticity map are features 
due to the lower resolution outside the highest refinement inner box ($\sim$2 kpc of box side).

The analysis presented below is centered on both the chemo-thermal and dynamical properties 
of the primordial gas. In this paper we do not analyze either the gas fragmentation process     
or the primordial gas clumps physical properties. This kind of study is postponed for 
a future work.

\subsection{Chemistry and Turbulence}
It is well known that below $\sim10^4$K the primordial gas can be cooled down only by 
molecular line emissions. In this sense the molecules are crucial for the formation of
star formation regions of cold and dense gas. 

The H$_2$ molecule is the most abundant molecule in primordial gas \citep{GP98}. Due to its symmetry
it has not a permanent dipolar moment and only quadrupolar rotational transitions 
are allowed \citep{Abgrall1982}. 

The main path for H$_2$ formation in primordial environments is the two steps channel \citep{Peebles1968}
\begin{equation}
\mathrm{e}^- + \mathrm{H} \rightarrow \mathrm{H}^- + h\nu,
\end{equation}
\begin{equation}
\mathrm{H}^- + \mathrm{H} \rightarrow \mathrm{H}_2 + \mathrm{e}^-.
\end{equation}
The H$_2$ molecule is able to drop the unperturbed primordial gas temperature till 
$T\sim 2-3\times10^2$K and till $T\ga100$K in perturbed regions with high electron 
fraction, e.g. in post shocked regions of turbulent gas with high Mach number or 
in relic HII regions.

On the other hand, the HD molecule, also present in primordial gas, has a permanent 
dipolar moment which allows faster rotational transitions. Furthermore, due to its 
higher reduced mass it has rotational energy levels lower than the H$_2$ levels,
thus reaching lower minimum gas temperatures due to rotational transitions \citep{FlowerRoueff1999}. 

The main path for HD production in primordial gas is based on the H$_2$ molecule (e.g. \citet{Flower2000}):
\begin{equation}
\mathrm{D}^+ + \mathrm{H}_2 \rightarrow \mathrm{HD} + \mathrm{H}^+.
\end{equation}
Due to the reasons mentioned above the HD molecule is able to cool the unperturbed 
primordial gas below 100K but under perturbed conditions the gas could reach the 
CMB temperature floor due to the HD cooling.
 
\subsubsection{H$_{2}$ and HD molecules}
It is interesting to understand how the gas dynamic affects 
the molecular formation process in primordial environments and which are the consequences 
of this highly non linear process.

Because H$_2$ production depends on the abundance of free electrons (and H$^-$)
and the HD production depends on the H$_2$ (and D$^+$) there are two paths to 
increase the molecular production rate: i) to increase the local gas density and
enhance the number of reactions, which can be a natural consequence in highly turbulent 
environments, or ii) to increase the number of free electrons by a given physical process, 
e.g. a ionization front. 

The first path can be the result of either a gravitational collapse process or a highly 
turbulent environment, as mentioned above. An example for the first case is the collapse 
of minihalos at high redshift \citep{McGreerBryan2008}; in this case the density enhancement due to the 
gravitational collapse produces enough molecules to reduce the primordial gas temperature 
to few $\times100$ K. An example for the second case can be the interaction of gas flows 
onto a (massive enough) DM halo; the flows can produce a turbulent enough environment 
where over-densities fill the interaction region \citep{Heitschetal2008}. 

In primordial environments the second path can be the result of post shock waves formed 
in the virialization process of massive enough DM halos. Despite the fact that previous works based 
on simplified unidimensional models have shown that post shock waves created through 
the baryonic collapse onto DM halos with masses M$<10^8$M$_\odot$ cannot create
enough free electrons in order to enhance the molecular production \citep{JohnsonBromm2006}, 
this is not completely true due to the real non uniform baryonic velocity field, 
i.e. for a given DM average circular velocity the gas content of the halo is able 
to reach velocities higher (and lower) than its average and then it could create 
shocks able to increase the free electron fraction even if the average circular 
velocity is not high enough. Furthermore, in the hierarchical 
paradigm of structure formation it is natural that the smallest halos interact with the 
biggest ones producing a merger process which can create shocks waves as well and 
consequently increase the free electron fraction due to a violent merger process.

\begin{figure*}
\centering
\includegraphics[width=15.5cm]{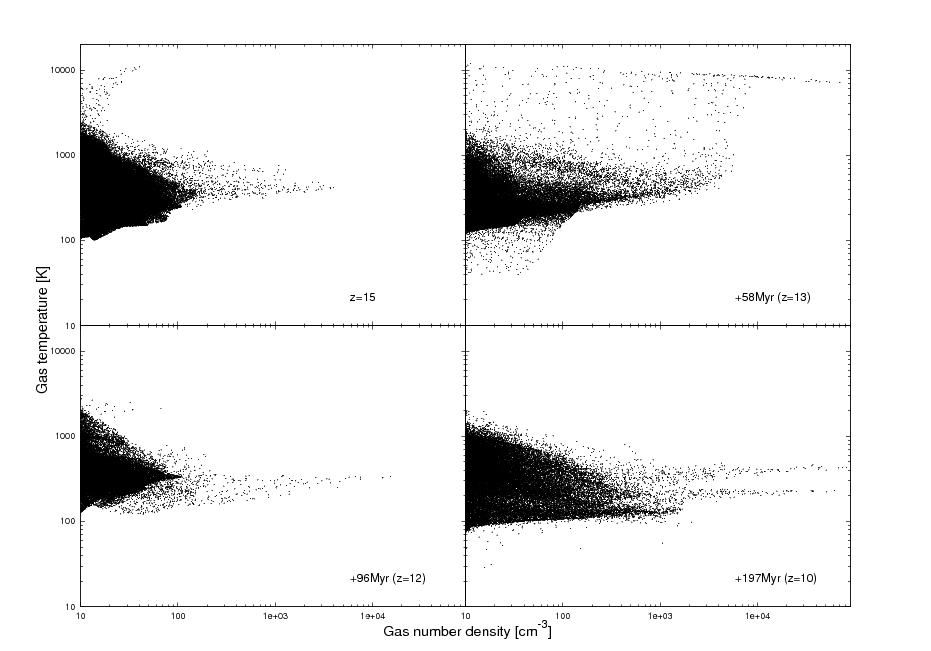}
\caption{Number density - gas temperature plane at z=15 (top left), 13 (top right), 12 (bottom left) and 10 (bottom right) for number densities $n>10$cm$^{-3}$. From the top right figure is clear how the merger process disturbs the densest main halo central region creating many very low temperature regions. These regions seem to be related to the post shock tail behind the spinning over-density. At later redshifts the temperature reaches values $T\la100$ K in the density range $10\la n/cm^{-3}\la10^3$ which can be associated to the HD molecular cooling of the turbulent main halo central region.}
\label{f1}
\end{figure*}

The number density vs. gas temperature plane of our simulation at z=15 (top left), 13 (top right), 12 
(bottom left) and 10 (bottom right) is shown in figure \ref{f1}. 
% Fifth minor comment
The figure shows number densities $n\ge10$cm$^{-3}$ corresponding to the densest gas of the simulation central regions. 
At z=15, before the merger, the gas temperature is above $\sim$400 K in the 
densest regions associated to the minihalos. After the merger, the gas develops very 
high temperature regions with $T\ga10^4$ K due to the violent merger process. At z=13 the high temperature gas has 
$T\la10^4$ K because of the hydrogen recombination lines cooling; This temperature is still enough in order 
to both dissociate the molecular coolants and enhance the ionization fraction. At this redshift, the high density regions, $n\ge\mathrm{few}\times10^3$ cm$^{-3}$, with high temperature, $T\la10^4$, are associated to the resultant hot-dense spinning over density going away the main halo central region, while the lower density hot regions correspond to gas expelled out by the explosive merger. At the same time, due to the enhanced ionization fraction the post shocked regions have 
had enough time to create HD molecules in order to drop the gas temperature till the CMB limit. 
% Fourth main point
(In our previous study of first galaxy formation in \citet{jpp2} the mass resolution was a factor of 8 lower than in this simulation. It is well known that in order to see the HD effect on the gas temperature it is needed a density threshold -\citet{McGreerBryan2008}-. Due to the higher resolution of this simulation here it is possible for the HD molecule dominate the gas cooling in some regions, including the formed gas clumps not analyzed in this paper.) 
At later times, when the dense-hot spinning over density is out the main halo central region (and certainly not captured in the bottom panels of figure \ref{f1}), a large amount of gas reaches temperatures $\sim100$ K in the density 
range $10\la n/\mathrm{cm}^{-3}\la10^3$: potential places for star formation. These final conditions are the result of a post-shocked turbulent evolution followed by gas accretion onto the central region. 
% First minor point

At the first glance our simulation seems contradict the result of \citet{ShchekinovVasiliev2006}. These authors found that in order to ionize the primordial gas in a halo merger processes, the halo mass should be $\ga10^7$ M$_{\odot}$, whereas in our simulation the merged halos have masses $\sim10^6$ M$_{\odot}$. The explanation for this apparent contradiction is that the result of \citet{ShchekinovVasiliev2006} are based on an isolated halo-halo interaction, whereas in our case the mini halos collide after to be accelerated by the few$\times10^7$ M$_{\odot}$ main halo, thus the merger is much violent than an isolated $\sim10^6$ M$_{\odot}$ halo-halo interaction.   

\begin{figure*}
\centering
\includegraphics[width=15.5cm]{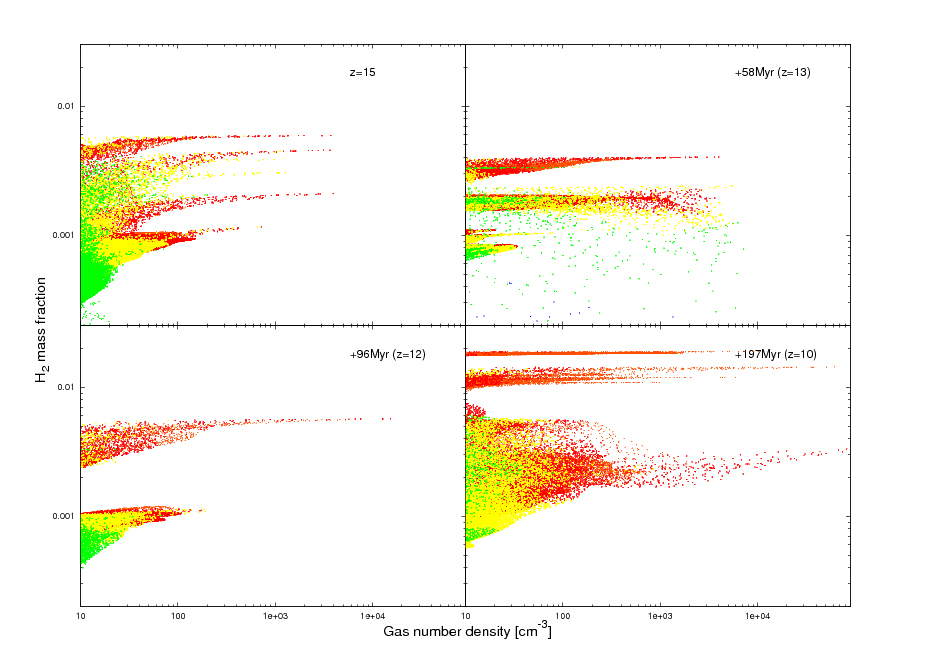}
\includegraphics[width=15.5cm]{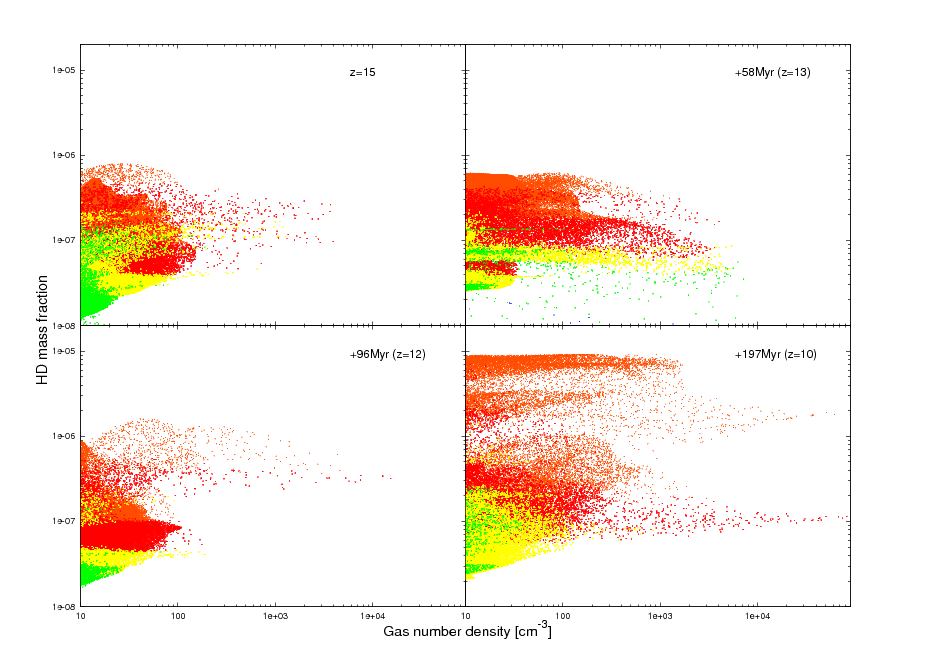}
\caption{Top: Number density - H$_2$ mass fraction plane at the same redshifts of figure \ref{f1}. In brown all grid cells with temperature in the range 100K to 300K, red  300K$<\mathrm{T}<$500K, yellow 500K$<\mathrm{T}<$1000K, green 1000K$<\mathrm{T}<$5000K and blue 5000K$<\mathrm{T}<$10000K. Due to the post shock conditions the H$_2$ molecule is able to reach very high mass fractions, $f_{H_2}\ga0.01$, which enables the gas to drop temperature $T\la200$K. Bottom: Same as top but for the HD molecule. Again, due to the post-shock conditions, the HD molecule can be produced very efficiently and it reaches $f_{HD}\la10^{-5}$. This high mass fraction allows the gas to reach temperatures $T\la100$K.}
\label{f2}
\end{figure*}

Figure \ref{f2} shows the number density - molecular mass fraction plane for both H$_2$ (top) 
and HD (bottom) molecules at the same four different redshift as in figure \ref{evolution1}. 
As mentioned above, before the merging there are a number of regions with temperature $T\ga400$ K 
and number densities $n\ga10^3$ cm$^{-3}$ associated with the minihalos (consistent with figure 
\ref{f1}). After the merger, both figures show that a lot of molecules are destroyed and the molecular 
mass fraction is reduced in a number of regions (top right panels). At the same time appear regions 
where molecules have continued working to reduce the gas temperature allowing the formation of denser 
zones. Approximately 40 Myr later (at z=12) the gas has developed very dense regions with $n\ga10^3$cm$^{-3}$.
Furthermore, there are new regions with $n\ga10^2$cm$^{-3}$ at lower molecular abundance. A hundred 
of million of years later there seems to be two populations of dense regions with $n\ga\mathrm{few}\times10^3$cm$^{-3}$: 
one with temperature below 300 K and H$_2$ mass fraction f$_{\mathrm{H_2}}\ga10^{-2}$ and HD mass fraction 
f$_{\mathrm{HD}}\la10^{-5}$ and another one with temperature above 300 K and f$_{\mathrm{H_2}}\ga2\times10^{-3}$ and 
f$_{\mathrm{HD}}\ga10^{-7}$. These two different regions can be the result of two different evolutions near 
the main halo central region. The coolest one seems to be the result of evolution onto a post shocked 
environment where the over-densities grow up in regions with an enhanced initial molecular mass fraction. 
The other over-densities seem to develop in regions not too much affected by the merger process 
presenting a molecular mass fraction one order of magnitude below \citep{McGreerBryan2008,JohnsonBromm2006}.

\subsubsection{Molecular cooling and gas velocity}
\label{MCGV}
The figures shown in the previous section suggest that the primordial gas dynamic triggers the formation 
of abundant H$_2$ and HD molecules through both the halo merging and the baryonic collapse 
process. Consequently, these molecules are able to cool the gas allowing it to reach 
$T\la$300K at density peaks, reaching temperatures even below 100K in some regions. 

\begin{figure*}
\centering
\includegraphics[width=15.5cm]{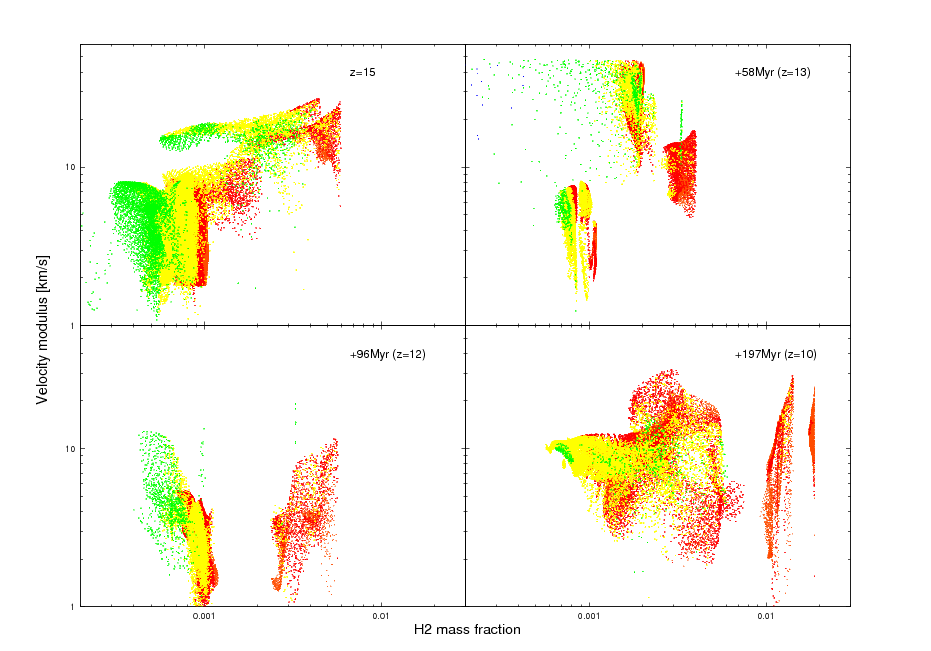}
\includegraphics[width=15.5cm]{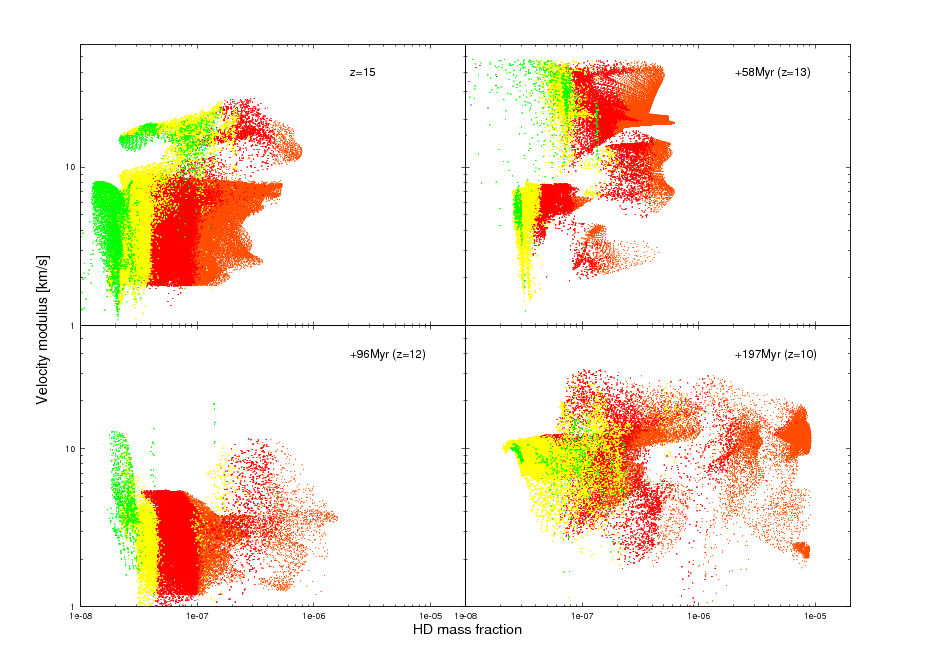}
\caption{Top: H$_2$ mass fraction - velocity modulus plane at z=15 (top left), z=13 (top right), z=12 (bottom left) and z=10 (bottom right). Bottom: Same as top but for HD molecule. The colors show the same range of temperature as in figure \ref{f2}. Both panels show how the merger process increases the gas temperature and creates very high velocity gas  which is able to create shocks and enhance the free electron fraction favoring the molecular formation in posts hock regions. After the merger, the gas of the main halo central region decreases its velocity but it is increased again by the baryonic accretion gas process at lower redshift. This accretion process is able to create high velocity gas waves which compress it and increase the molecular mass fraction.}
\label{f3}
\end{figure*}

Based on this dynamical dependence of molecular formation it is interesting to know how the
velocity modulus/Mach number is related with the number density, gas temperature and molecular 
mass fractions. Figure \ref{f3} shows the molecular mass fraction - velocity modulus planes 
for both the H$_2$ (top) and the HD (bottom) molecule at the same redshifts of previous figures. 
% First main comment (label physical/comoving)
The velocity modulus is computed as the value of 
$[(v_{\mathrm{grid,x}}-v_{\mathrm{av,x}})^2+(v_{\mathrm{grid,y}}-v_{\mathrm{av,y}})^2+(v_{\mathrm{grid,z}}-v_{\mathrm{av,z}})^2]^{1/2}$ 
for each grid; here $v_{\mathrm{av,i}}$ is the average gas velocity inside the analyzed box of $\approx22$ comoving kpc, i.e. $(\sum v_{\mathrm{grid,i}})/N_{\mathrm{grid}}$, 
where $i=x,y,z$; $v_{\mathrm{grid,i}}$ is the local grid velocity and $N_{\mathrm{grid}}$ is the total number of grids inside the highest resolution volume.
We emphasize that the analyzed volume is not the entire 1 Mpc$^3$ simulation comoving box but the innermost highest resolution 
volume of $\sim2$ proper kpc at z=10 ($\sim1.4$ proper kpc at z=15). Keeping this in mind, the gas velocity definition of $\vec{v}_{\mathrm{gas}}=\vec{v}_{\mathrm{grid}}-\vec{v}_{\mathrm{av}}$
has all sense because it represents the local grid velocity without take into account the gas average movement of the 
analyzed volume.
% end
(The computed rms physical velocity through the evolution are 8.18 km/s at z=15, 7.06 km/s at z=13, 6.21 km/s at 
z=12 and 4.18 km/s at z=10.) Despite of the main halo circular velocity $V_{\mathrm{circ}}\approx14.6$ 
km/s, the main halo central regions develop gas velocities above 22 km/s, enough to enhance the ionization 
fraction and produce shock waves \footnote{For comparison the average circular velocity of a $10^8$M$_\odot$ 
DM halo collapsed at $z\approx10$ is $v_{\mathrm{circ}}\approx21.8$ km/s.} or at least enough to 
compress significantly the primordial gas and accelerate the molecular formation process. At 
z=13, about 40 Myr after the merger process, the gas reaches velocities near 60 km/s 
evidencing the violence of the mini halo interaction. The interaction is able to enhance the 
temperature till $\sim10^4$ K, slightly increase the free electron fraction and trigger an efficient 
molecular formation process.  

\begin{figure*}
\centering
\includegraphics[width=15.5cm]{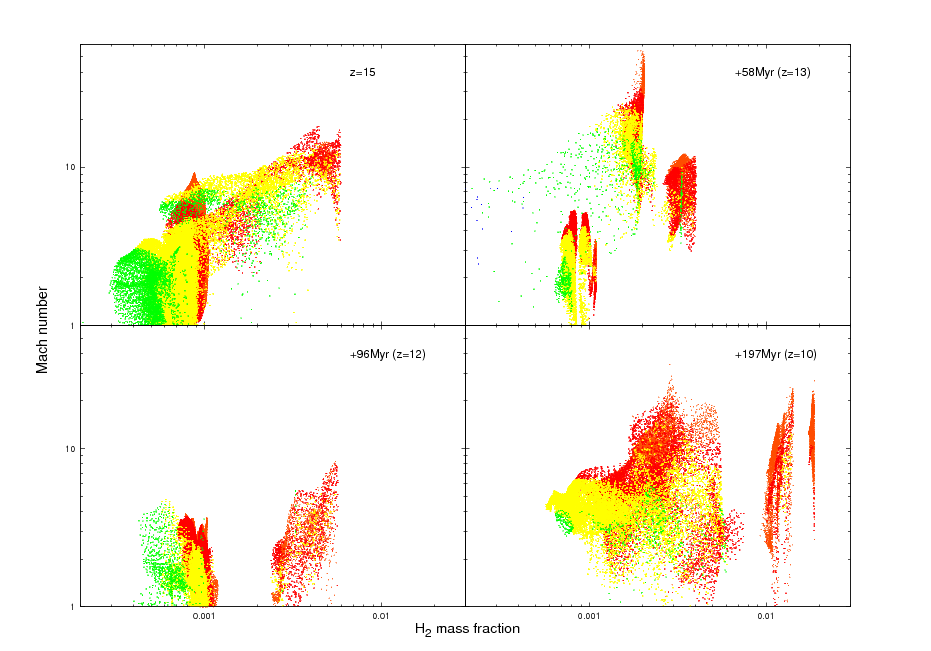}
\includegraphics[width=15.5cm]{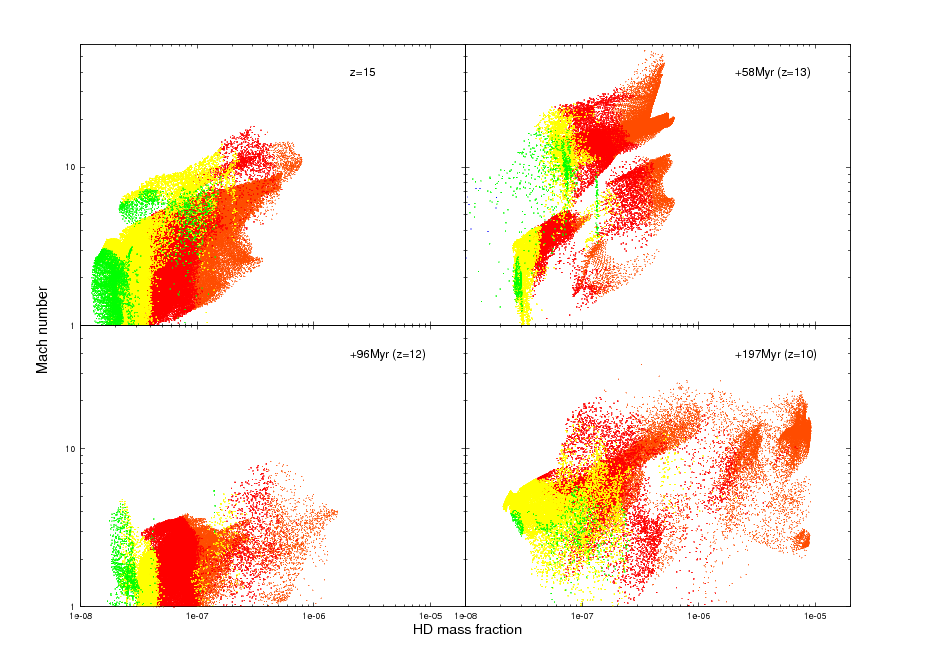}
\caption{Top: H$_2$ mass fraction - Mach number plane at the same redshifts as previous figure. Bottom: Same as top but for HD molecule. The colors show the same range of temperature as in figure \ref{f2}. Both planes shows that Mach numbers above 10 are reached in regions with both high molecular abundances and low temperatures. In general the coolest regions tend to show the highest Mach numbers which relates turbulent regions with high molecular abundances evidencing the effect of merger and supersonic turbulence on the primordial gas, i.e to increase the molecular mass fraction inducing efficient cooling.}
\label{f4}
\end{figure*}

The molecular mass fraction - Mach number planes for both H$_2$ (top) and HD (bottom)
molecules at the same redshifts as in previous figures are shown in figure \ref{f4}. The
Mach numbers were computed based on the velocity modulus defined above.
From these figures we can see that through the evolution almost all primordial gas with 
$n>10$ cm$^{-3}$ develops super sonic waves. After the merger there are regions with 
Mach number near 50 evidencing the violence of the process again. At later times the 
Mach numbers (and velocities) decay at the main halo central region to increase again 
at z=10, showing how after the merger the gas relaxes and reaches Mach numbers below 10 at 
the densest regions but it increase again promoted by the baryonic accretion process. 
These two figures show that the coolest gas has the highest molecular mass fraction 
but interestingly this coolest gas occupies the high velocity/Mach number regions as a proof 
that the highly supersonic turbulent gas is composed by a high molecular mass fraction.

\begin{figure*}
\centering
\includegraphics[width=15.5cm]{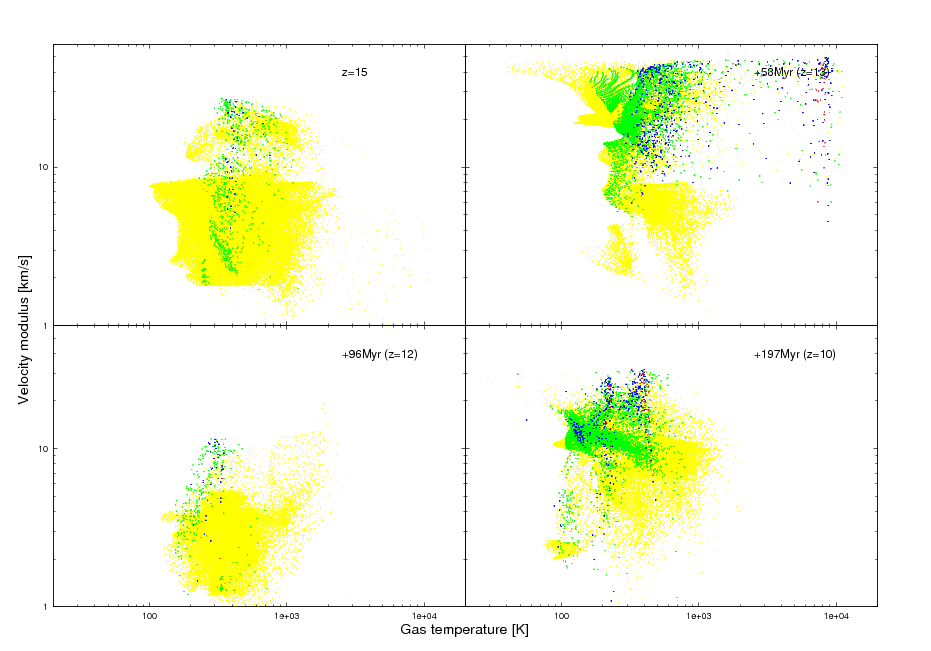}
\includegraphics[width=15.5cm]{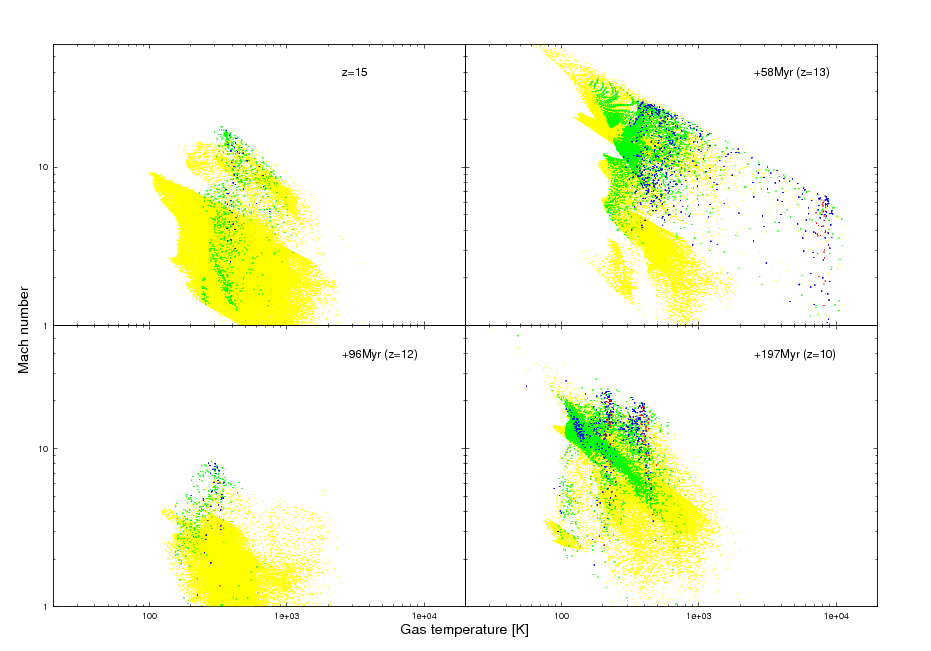}
\caption{Top: Gas temperature - velocity modulus plane for $n>10$ cm$^{-3}$. In yellow $10^1\le n/\mathrm{cm}^{-3}< 10^2$; in green $10^2\le n/\mathrm{cm}^{-3}<10^3$; in blue $10^3\le n/\mathrm{cm}^{-3}<10^4$ and in red $10^4\le n/\mathrm{cm}^{-3}<10^5$. Top: The figure show that high density gas with $n\ga100$ cm$^{-3}$ is able to reach velocities above $v_{\mathrm{rms}}\approx20$ km/s, high enough to enhance the free electron number density and compress the gas increasing the molecular mass fraction. Bottom: Same as top but for the gas temperature - Mach number relation. High density gas with temperature below $10^3$ K shows supersonic velocities with the highest Mach numbers associated to the lowest temperature regions. The high density regions with $n>100$ cm$^{-3}$ clearly present super-sonic turbulence. This feature relates low temperature - high Mach number with high density regions showing a clear relation among super sonic turbulence, molecular cooling and high density regions.}
\label{f5}
\end{figure*}

At the moment we have a picture in which the mergers and the gas collapse process onto 
the main halo triggers efficient molecular formation in high velocity/supersonic gas regions 
allowing the gas to reach temperatures even below 100 K. Figure \ref{f5} shows the gas 
temperature - velocity modulus plane (top) and the gas temperature - Mach number plane (bottom), 
for number density above 10 cm$^{-3}$, both at the same redshift as before. 
% Sixth minor comment
%The different colors indicate: in yellow $10^1\le n/cm^{-3}< 10^2$; in green $10^2\le n/cm^{-3}<10^3$;
%in blue $10^3\le n/cm^{-3}<10^4$ and in in yellow $10^4\le n/cm^{-3}<10^5$.
From this figure
it is clear how the merger process creates high density regions with temperature $\sim10^4$ K
due to its violence. At lower redshifts this figure shows that the high density regions with 
$n\ga10^2$ cm$^{-3}$ are associated with both high Mach numbers and high velocities 
suggesting that the collapsed gas converging in the main halo central region is compressed 
at velocities 10 km/s $\la v_{\mathrm{rms}}\la$ 30 km/s, which can help to increase both the 
molecular abundances and consequently the gas cooling. In this way the mergers and the baryonic 
matter collapse process would favor the formation of cold - dense primordial gas regions and 
creates potential primordial star formation clouds. 
% Fourth main comment: In this sim the densities are higher, then we can see the HD effect.

\subsection{Probability Distribution Function (PDF)}
In the hydrodynamical context, the density PDF $p$ gives the probability to find a fluid 
element with a density between $\rho$ and $\rho + \mathrm{d}\rho$ inside the analyzed volume. 
In other words it quantifies the fraction of the gas in a given range of density as
$\int_\rho^{\rho+\mathrm{d}\rho} p(\rho') \mathrm{d}{\rho}'$. 
 
\citet{VazquezSemadeni1994} shows that in the high Mach number limit, without viscosity
and neglecting the gravitational term, the normalized Euler equations for hydrodynamics 
describe a scale invariant pressureless fluid, i.e. the probability to create a relative 
density fluctuation in a given region is independent of the local density. As explained 
in \citet{Slyz+2005}, if we assume that in a full developed turbulent fluid the density 
field is a random variable and the event of successive density increments are independent, 
then the central limit theorem states that the density PDF is lognormal\footnote{The PDF 
is lognormal because the random variable density is the product of independent random 
variables instead the sum of random variables, in the last case the PDF should be a normal 
distribution.}.

The argument exposed above has been extensively studied using hydrodynamical simulations of 
isothermal supersonic turbulence without gravity (e.g. \citet{Kritsuk+07,Federrath+08}). These 
simulations have shown that the density PDF is well represented by a lognormal distribution:
\begin{equation}
p(\ln \rho)\mathrm{d}\ln \rho = \frac{1}{\sqrt{2\pi\sigma^2}}e^{-\frac{1}{2}\left(\frac{ \ln\rho-\overline{\ln\rho} }{ \sigma }\right)^2}\mathrm{d}\ln\rho,
\end{equation}
where 
\begin{equation}
\overline{\ln\rho}=-\frac{\sigma^2}{2},
\end{equation}
with $\sigma$ the standard deviation of the logarithm of the density. For a formal proof
of the lognormal density PDF distribution based on \citet{Pope+Ching93} see \citet{Padoan+Nordlund1999}.

The lognormality of the PDF is valid for isothermal turbulent fluids without gravity but
under realistic conditions the asumption of isothermality is not well justified due to 
the cooling/heating processes triggered by both the chemistry and the gas dynamic. 
Furthermore, in less than a dynamical time the selfgravity becomes unavoidable at the high density regions in a realistic scenario. Therefore, we do not expect {\em a priori} to find a perfect 
log-normal density PDF in our non-isothermal simulation with self gravity. 

\begin{figure}
\centering
\includegraphics[width=9.5cm]{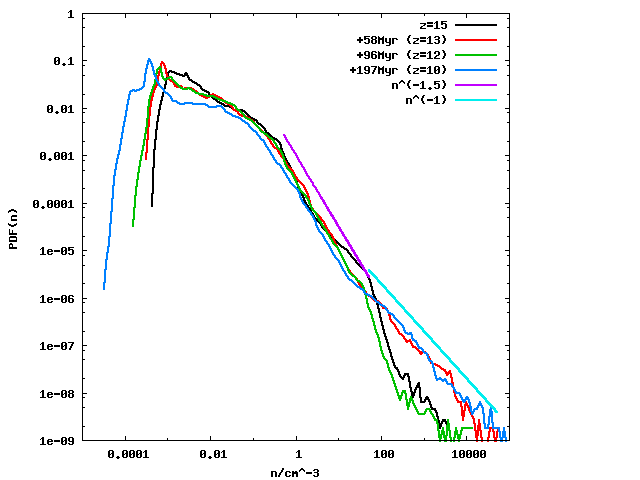}
\includegraphics[width=9.5cm]{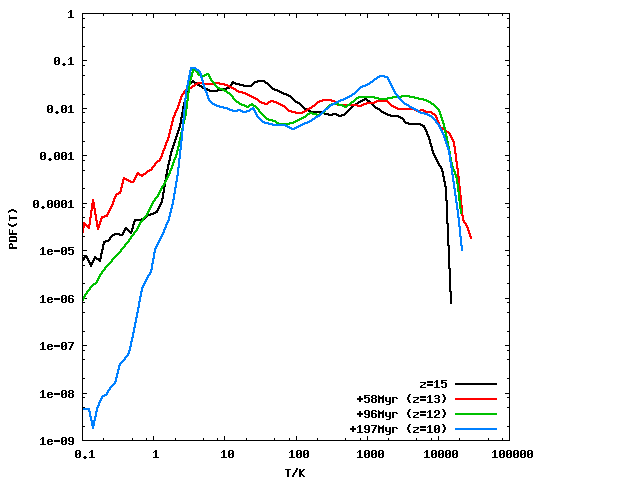}
\caption{Top: number density probability distribution function for different redshifts. In black: z=15, red z=13, green z=12 and blue z=10. This figure shows that the density PDF has not a symmetric log-normal distribution and a high percentage of the analyzed volume corresponds to low density, $n\la0.1$ cm$^{-3}$, regions. As explained in the text, the high density tail covers two regimes: the first one for $0.5\la n/\mathrm{cm}^{-3}\la50$ shows a $n^{-3/2}$ behavior (purple line) associated to a nearly isothermal collapse process; the second one, at densities $n\ga50$ cm$^{-3}$ follows a $n^{-1}$ behavior (light blue line) associated to the disk structures of collapsed gas clumps. Bottom: Same as top but for the gas temperature. At z=10, the temperature PDF shows two clear peaks consistent with a two phase fluid of low temperature regions coexisting with high temperature regions. These two peaks are not so clear at z$>$13 and develop through the gas collapse process.}
\label{f6}
\end{figure}

Figure \ref{f6} shows the PDF of both the gas density (top) and the gas temperature (bottom) 
for redshifts from 15 to 10. The PDFs (for density, temperature and velocity) were computed 
as the histogram of the given quantity inside equal-size bins in logarithmic scale. The 
density PDF shows a clear deviation from a log-normal distribution which as mentioned above 
is a consequence of the non isothermal evolution with self gravity. 

It is interesting to note how the high density tail of the PDF evolves with redshift; it takes 
higher values at lower redshift showing how the gas develops high density ($n\ga10^2$cm$^{-3}$) 
regions through the collapse process onto the main DM halo; this high density regions are less than
$\sim 5$\% of the volume at all redshifts. 

The features of the PDF high density tail have been recently studied by \citet{Kritsuk+11} 
in isothermal supersonic turbulence simulations with selfgravity in a periodic box of 5 pc size. 
Despite of the difference in scale among this two simulations, the PDFs are roughly similar in shape. 
The yellow line in figure \ref{f6} shows a $p(n)\propto n^{-1.5}$ whereas they found a $n^{-1.695}$ 
power law. This behavior is explained as the result of the first stage of a nearby isothermal 
gravitational collapse, as mentioned in their paper. The blue line shows a $p(n)\propto n^{-1}$ 
(they found a $p(n)\propto n^{-0.999}$) which is explained as the region in which their gas cores 
formed by gravitational collapse are rotationally stable. Interestingly, the same explanations could 
be valid in our simulation because through the gas collapse process onto the main DM halo the gas
develops over-densities at the turbulent super sonic regions which eventually end as primordial 
gas clumps of few$\times$100 K due to molecular cooling. These over-densities show intrinsic 
rotation and develop gravitationally supported disk structures at lower redshifts.

The average density inside the analyzed box for all redshift of figure \ref{f6} is $\bar{n}\approx (1-2)\times10^{-2}$ cm$^{-3}$. 
The density PDF shows a peak at very low densities (few$\times10^{-4}$cm$^{-3}$) instead that  
at the average density like in the lognormal isothermal distribution. This peak is easily explained 
in the collapsing scenario: through the collapse process, the baryonic matter piles up near the 
main halo central region. While in this region the gas develops over-densities in a tiny percentage
of the volume, the rest of the volume is dominated by low density regions due to the converging 
gas flow onto the main halo. This dynamical behavior creates more low density regions at lower 
redshift as shown by the density PDF and favors the formation of new high density regions. 

The temperature PDF in figure \ref{f6} shows two clear peaks at z=10, one at $\sim$ few K and 
another one at $\sim$ few $\times10^3$ K. These two peaks are not easily recognized at $z>13$ and
they appear at lower redshift as a consequence of the baryonic matter accretion process. The two
peaks feature suggests the existence of two preferred gas states of high and low temperature co-existing inside the analyzed volume. Again, in the collapsing scenario the low temperature gas 
can be associated to decreasing (expanded and very low) density regions -which are not in equilibrium 
with the CMB photons- due to the infalling gas flow onto the main halo central region. These regions 
contain the gas inside the voids surrounded the high density structures (this is the reason why they do not appear in the figures shown above which show gas with $n>10$ cm$^{-3}$). On the other hand, 
the high temperature regions can be associated with heated gas infalling onto the DM halo through the 
virialization process. Between these two temperature peaks is located the $\sim$ few $\times10^2$ K 
gas, where the molecular coolants are working allowing the formation of high density regions at lower 
redshift. Of course the molecular cooling region is a tiny percentage of the total $\sim$ few 
$\times10^2$ K range which should be dominated by thermally unstable gas.    

\begin{figure}
\centering
\includegraphics[width=9.5cm]{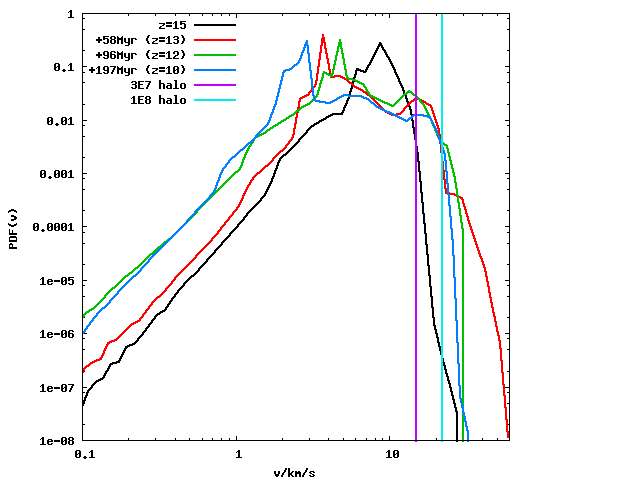}
\caption{Gas velocity PDF at different redshifts. In black: z=15, red z=13, green z=12 and blue z=10. The vertical lines show the main halo average circular velocity of 14.6 km/s (in purple) and the average circular velocity of a $10^8$ M$_{\odot}$ halo (in light blue): 21.8 km/s. The time evolution of the PDF shows how the minihalo merger process creates regions with velocities near 60 km/s. After that the maximum gas velocity decreases to values near 30 km/s mainly supported by the gas accretion process onto the main halo.}
\label{f7}
\end{figure}

Because we have shown that the gas dynamic is related to the cooling process in primordial 
environments, it is interesting to study the mean gas velocity PDF of the system. Figure \ref{f7}
shows the mean gas velocity PDF for the same redshifts analyzed in previous sections. The two 
vertical lines are the the main halo average circular velocity (14.6 km/s in purple) and the 
average circular velocity of a $10^8$ M$_{\odot}$ halo (21.8 km/s in light blue). At z=15, before 
the minihalo merger the gas velocity is below $\sim$20 km/s, a velocity too low in order to create 
shock waves. At z=13, after the merger, the gas develops regions with velocities near 60 km/s and 
it shows a longer high velocity tail. At this redshift, above 1\% of the gas has mean velocities 
above 21.8 km/s. At lower redshifts this percentage slightly increases but does not reach the 2\% 
of the volume. Furthermore, below z=13 the maximum velocity slightly surpass the 30 km/s, showing 
that the very high velocities triggered by the merger are replaced by moderately high velocities 
promoted by the gas accretion process on the main halo.

% Third main comment: The PDF are useful for idealized hydro simulations of star formation.
\subsection{Structure Functions}
\begin{figure}
\centering
\includegraphics[width=9.5cm]{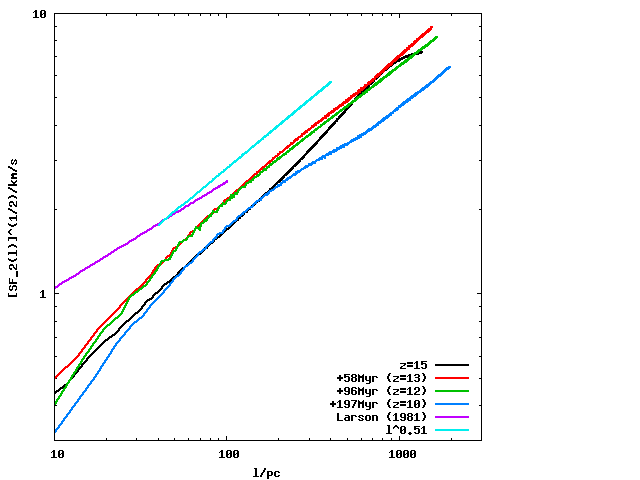}
\caption{Root square of the second order longitudinal velocity structure function for different redshifts. In black: z=15, red z=13, green z=12 and blue z=10. The light blue line shows a $S_2(\ell)^{1/2}\propto\ell^{0.51}$ which is a good approximation for all curves in the range $50$pc$\le\ell\le200$pc. This range of scales roughly coincides with the molecular cooling (energy dissipation) scale and gas accretion (energy injection) scale, in other words the power law is satisfied roughly in the inertial range of the system. The purple line correspond to the size-velocity relation found by \citet{Larson81} in local galactic molecular clouds.}
\label{f8}
\end{figure}

In order to characterize the turbulence, we computed the second order velocity
structure function. The velocity structure function of order $p$ is defined
as:
\begin{equation}
S_p(\ell)=\langle | v({\it \bf x}+{\bf \ell})-v({\it \bf x})|^p\rangle \propto \ell^{\zeta(p)},
\label{sf}
\end{equation}
where the velocity component $v$ is parallel (longitudinal structure function) or
perpendicular (transversal structure function) to the vector $\ell$. The spatial 
average is over all values of the position ${\it \bf x}$ and ${\zeta(p)}$ is the 
exponent of a power law fit to the structure function. The second order longitudinal 
structure function averaged over the central region of the simulation from z=15 to z=10
is plotted in figure \ref{f8}. It is well approximated by a power law, $\ell^{1.02}$, 
in the range of scales $\ell=50-200$ pc (light blue line) for the given range of redshifts. 

Taking into account that after the merger the dynamics of the system is supported by the matter 
accretion process at scales of $\mathrm{L}_{max}\ga500$ pc, and the molecular cooling process 
is efficient (in high density regions) at scales $\mathrm{L}_{mim}\la\mathrm{few}\times10$ 
pc, the range in which the power law is valid corresponds roughly to the inertial range 
of the system. 

In order to relate this result with the observed properties of local galactic star-forming regions, where the turbulence is driven mainly by SNe's shock waves, figure 
\ref{f8} actually shows the square root of the second order structure function next 
to the velocity scaling law found in galactic molecular clouds \citep{Larson79,Larson81}, 
which follow the relation $S_2(\ell)\propto\ell^{0.76}$. Interestingly, on the scale of 
approximately 100 pc, the velocity dispersion in the primordial gas of our simulation is 
comparable to that of the molecular gas in our galaxy. However, due to the larger temperature 
in the primordial gas, the Mach number of the turbulence is a few times smaller than in 
nearby molecular clouds at the same scale.

\subsection{Fourier Analysis}
\subsubsection{Power Spectrum (PS)}
After we have shown that supersonic turbulence, triggered by both merger processes and gas 
accretion, promotes molecular formation and gas cooling in primordial environments, it is 
interesting to study how the kinetic gas energy is distributed in their different Fourier
modes.

The kinetic energy PS is defined as:
\begin{equation}
\widehat{E}_{k}=\frac{1}{2}(4\pi k^2 \hat{v}_{k}\cdot\hat{v}_{k}^{*}), 
\end{equation}
where $\hat{v}_{k}$ is the gas velocity Fourier transform, $\hat{v}_{k}^*$ its complex 
conjugate and $k\equiv|\vec{k}|$ the modulus of the wave number. $\widehat{E}_{k}$ represents the 
kinetic energy associated to Fourier modes of wave number between $k$ and $k+\mathrm{d} k$. The 
kinetic energy spectrum is useful to distinguish between subsonic Kolmogorov turbulence, with 
$\widehat{E}_{k}\propto k^{-5/3}$, and shock dominated Burgers turbulence, with $\widehat{E}_{k}\propto k^{-2}$. 

\begin{figure}
\centering
\includegraphics[width=9.5cm]{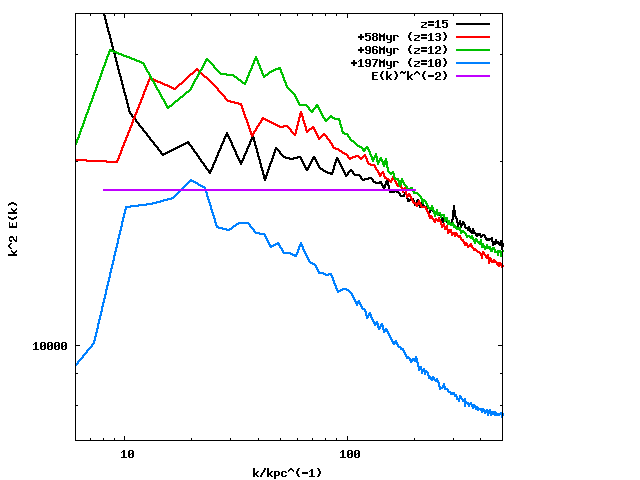}
\caption{Velocity power spectra for different redshifts. In black: z=15, red z=13, green z=12 and blue z=10. The purple line correspond to a power spectra $\widehat{E}_k\propto k^{-2}$. This line is a good approximation at z=15. At lower redshift the power spectra is a bit steeper and it is close to a value of $2.2$. The near to Burgers behavior is consistent with a supersonic environment.}
\label{f9}
\end{figure}

The energy PS for our analyzed volume is shown in figure \ref{f9}. This figure shows the
kinetic energy PS from z=15 to z=10. The purple line shows an $\widehat{E}_{k}\propto k^{-2}$
Burgers PS. It is clear that at z=15 (black line) the gas energy spectrum follows a nearly 
Burgers behavior from $k\approx10$ kpc$^{-1}$ (physical scale $\approx$600 pc) to $k\approx100$ 
kpc$^{-1}$ (physical scale $\approx$60 pc). Actually, the power law exponent is a bit steeper at 
this redshift. At lower redshifts the power law exponent decreases and it reaches values near -2.2 
at z=10. This result is in good agreement with the second order longitudinal velocity structure 
function shown in the previous section: for a $\hat{E}_k\propto k^{-n}$ the $S_2(\ell)\propto \ell^{n-1}$, 
which for $n\approx2$ gives approximately $S_2(\ell)\propto \ell^{1}$ as shown in figure \ref{f8}. 

The close-to-Burgers power law indicates that the turbulent motions triggered by both the merger
process and the matter accretion process onto the main halo produce supersonic velocities. 
This result is in agreement with our results of subsection \ref{MCGV}, where we have shown that 
the primordial gas develops supersonic $\mathcal{M}\ga10$ motions through the collapse process 
in high density regions, near the main halo central region.

It is interesting to note that at high wave numbers the PS does not decay as fast as in 
supersonic isothermal simulations without selfgravity \citep{Kritsuk+07}. This feature can be 
understood because at small scales -of order few $\times 10$ pc in high density regions- the 
gravitational potential energy is converted into kinetic energy which is distributed among 
the longitudinal and transverse velocity modes as we will show in the next subsection. 
In this sense, in the same way as the large scale accretion process drives the turbulent 
and shocked environment at the main halo central region, at small scales the gravity is 
able to increase the kinetic energy through a turbulent energy cascade which favors the 
formation of gravitationally bounded cold dense clumps \citep{Slyz+2005,Federrath+11}.

As mentioned in subsection \ref{MCGV}, the supersonic turbulence and the molecular cooling seem
to work togheter developing high density regions: the large scale collapse triggers the 
formation of supersonic turbulent regions where both the H$_2$ and the HD are formed efficiently 
cooling down the gas and allowing gravity to dominate at few $\times 10$ pc scales. These 
phenomena could rise the velocity powers at small scales compared with the non selfgravity case 
as shown in figure \ref{f9}.
% Third main point
One interesting thing to take into account for future idealized primordial star formation simulations, e.g. \citet{Clarketal2011}, is the PS slope found in this simulation. Our $\approx-2$ slope is different to the assumed -4 slope in \citet{Clarketal2011}. This difference could be not negligible because it could imply more fragmentation at small scales, certainly an interesting point to study.
\subsubsection{Solenoidal v/s Longitudinal Modes}
The Helmholtz decomposition of the velocity field allows us to study the kinetic energy
content in compressional modes ($\hat{v}_{c,k}$ with $\nabla\times \hat{v}_{c,k}=0$) and
in solenoidal modes ($\hat{v}_{s,k}$ with $\nabla\cdot \hat{v}_{s,k}=0$). In Fourier space
these two components are defined as:
\begin{equation}
\hat{v}_{c,k} = (\hat{v}_{k}\cdot\vec{k})\vec{k}/k^2. 
\end{equation}
\begin{equation}
\hat{v}_{s,k} = \hat{v}_{k}-(\hat{v}_{k}\cdot\vec{k})\vec{k}/k^2. 
\end{equation}

\begin{figure}
\centering
\includegraphics[width=9.5cm]{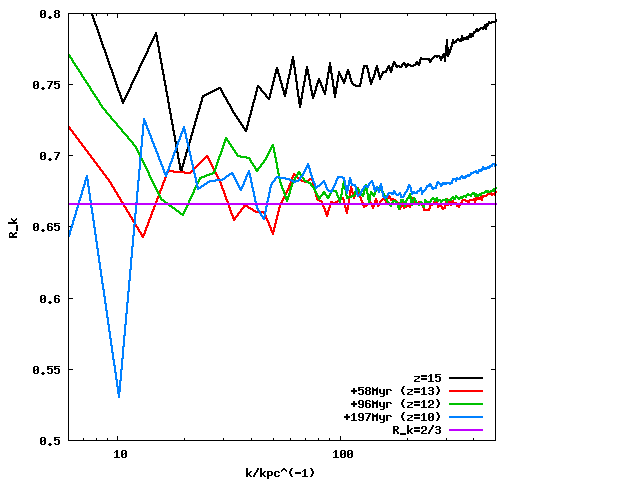}
\caption{Solenoidal to total energy ratio $R_k$ for different redshifts. In black: z=15, red z=13, green z=12 and blue z=10. For a long range of $k$ the energy ratio takes values $0.65\la R_k\la0.7$. These values are very close to the $R_k\approx 2/3$ found in high resolution hydrodynamical simulations and is a natural consequence of a system with fully developed turbulence.}  
\label{f10}
\end{figure}

Figure \ref{f10} shows the solenoidal to total kinetic energy ratio
\begin{equation}
R_{k}=\frac{\frac{1}{2}(4\pi k^2 \hat{v}_{s,k}\cdot\hat{v}_{s,k}^{*})}{\frac{1}{2}(4\pi k^2 \hat{v}_{k}\cdot\hat{v}_{k}^{*})} 
\end{equation}
from z=15 to z=10. For redshifts below 15 and for all wave numbers shown in the 
plot the solenoidal modes contain more than 50\% of the total kinetic energy, i.e. 
$R_k>0.5$. Furthermore, it is very interesting to note that above $k\approx30$kpc$^{-1}$
(physical scales below $\sim200$pc) $R_k$ naturally takes values between 0.65 and 0.70, 
near a 2/3 distribution found in previous works \citep{Federrath+11} which arises naturally 
in a three dimensional space where for a given compressional direction there are two
rotational directions. It seems that at z=15 the gas has not distributed the energy following 
an equipartition rule. May be it needs the merger and the gas accretion onto the main 
halo in order to redistribute its energy to get the 2/3 value, in other words, the gas needs 
to pass through chaotic processes to achieve equipartition of energy.

Under these result we can say that both the minihalo merger process and the large scale 
gravitational collapse process triggered by the main DM halo creates a supersonic turbulent 
cascade which naturally distribute the kinetic energy in a ratio $\hat{E}_c/\hat{E}_s\approx 0.5$, 
where $\hat{E}_s$ and $\hat{E}_c$ are the kinetic energy in solenoidal and compressional modes, 
respectively.
% Third main comment: Both the power spectrum and the energy partition are useful for idealized hydro simulations of star formation. In particular Clark et al. use a higher spectral index which could avoid the formation of stars at smaller scales. 
\section{Discussion}
In this paper we have shown that the mini halo merger process inside a $\approx3\times10^7$ 
M$_{\odot}$ halo is able to form shocks waves and supersonic turbulence in primordial gas. 
Furthermore, we have shown that the gas accretion process onto the main DM halo creates more 
supersonic turbulence and maintains the turbulence formed in the merger process. It is interesting 
that despite of the halo mass $M<10^8$ M$_{\odot}$ (the limit suggested by \citet{JohnsonBromm2006}) 
it is possible to find gas elements with enough velocity to create shocks (due to both the mergers 
and the gas accretion). As mentioned above, it is not surprising if we think that the 
estimation of \citet{JohnsonBromm2006} is based on the average halo circular velocity and do 
not take into account the velocity distribution of a more realistic system. The inclusion of 
a realistic velocity distribution would let the gas to reach velocities higher (and lower) 
than its average circular velocity and creates shocked regions. This could be possible for 
halos with masses well below $M<10^8$M$_{\odot}$ as has been previously shown by \citet{Greifetal2008} and by this work. But here we have also demonstrated, for the first time, that even very low mass halos ($M \sim 10^6 M_{\odot}$), because of the presence of a $10^7 M_{\odot}$ halo, can be perturbed enough to produce turbulence and shocks inside themselves, thus triggering more efficient cooling than without the merging process. 
% First minor point
The previous fact does not contradict the results of \citet{ShchekinovVasiliev2006} and \citet{VasilievShchekinov2008} whom by one dimensional simulations shown that in order to ionize the primordial gas in a halo merger process, the halo mass should be $\ga10^7$ M$_{\odot}$. Despite in our case the merged halos have masses $\sim10^6$ M$_{\odot}$, they are accelerated by a $\sim10^7$ M$_{\odot}$ halo and they can reach enough velocity to produce the gas ionization. 

The supersonic turbulence formed through this process favors the molecular formation by two main 
channels: i) the strong shocks of velocities above 22 km/s (due to both gas accretion and mini halo merger) are able to enhance the ionization 
fraction which catalyzes both the H$_2$ and the HD formation increasing their abundances in post 
shocked regions with the consequent gas cooling, and ii) the supersonic turbulence compresses the 
gas and increases the local gas density favoring molecular formation, which in turn enhances the lost 
of energy rate in the over-densities. The first channel is possible because the electrons 
recombine in time scales of $t_{\mathrm{rec}}\ga 10$ Myr while the molecular coolants are 
formed in scales of $t_{\mathrm{mol}}\la$ few Myr, under post shock conditions in these 
environments. The second channel works on time scales $t_{\mathrm{mol}}\la 10$ Myr for over densities 
of $n\sim10^2$ cm$^{-3}$. The combination of these two process, triggered by both halo merger and gas accretion, facilitates  
the gas to drop its temperature and to form high density regions of $n\ga10^4$cm$^{-3}$ with 
temperatures below 300 K, in other words: it allows to create cold-dense gas regions which 
are potential places for star formation. The gas temperature reaches values below 100 K at 
regions with $50\la n/\mathrm{cm}^{-3}\la10^3$ where the HD molecule seems to be a more 
efficient coolant than the H$_2$ molecule, at higher densities the gas cooling again is 
dominated by H$_2$ allowing temperatures $\sim (2-3)\times10^2$ K.

The combined effect of gas dynamic (merger and accretion) and gas chemistry (molecular 
formation and cooling) is able to produce over-densities with high velocity and high Mach number. The highly supersonic densest regions of the main halo central region develop a density PDF with interesting features. In the low density region it shows a peak below the average density inside 
the analyzed volume. This peak is formed because through the gas accretion process a large percentage 
of the matter is piled up in the central region. While the accretion flows converge near the halo 
center, the regions surrounding it lose gas continuously increasing the fraction of low density 
volume at lower redshift. 

At the other extreme, the high density PDF shows two clear power law region. The first at $0.5\la n/\mathrm{cm}^{-3}\la50$ 
follows a PDF$\propto n^{-1.5}$ which is explained as the region dominated by nearby isothermal
over-densities (the over-densities have to scale as $n\propto r^{-2}$ in order to reproduce the 
-3/2 exponent) before collapse. At higher densities, $n\ga 50$ cm$^{-3}$, the PDF tail shows how 
the collapse process at large scales feeds the central region and produces the high density structures. 
The break down of the -3/2 slope shows the point where the locally collapsed over-densities start to 
pile up the accreted material and develop rotationally supported structures as a natural consequence 
of angular momentum conservation. Very interestingly the density PDF computed from our cosmological simulation has similar features to the PDF found in idealized star formation simulations, e.g. \citet{Kritsuk+11}.

We emphasize that the density, temperature and velocity PDF can be used in order to characterize the gas conditions in future idealized primordial star formation simulations inside of similar high redshift environments .

The previous evidence of supersonic environments is further supported by our Fourier analysis. The kinetic energy spectrum of the turbulent gas follows a nearly Burgers behavior $\propto k^{-2}$, a bit steeper 
at lower redshift reaching a minimum exponent of $\approx$-2.2. This result is also supported by the second order longitudinal structure function exponent $\zeta_2(\ell)=\ell^{1.02}$. The power spectrum 
slope confirms that merger followed by the inhomogeneous accretion process onto the main halo develops a supersonic turbulent gas with shocks through the halo virialization process. An interesting feature of 
this spectrum is that toward small scales it does not decay as fast as in isothermal simulations with 
no selfgravity. This feature could be explained by taking into account the selfgravity of the densest 
regions: At the beginning, the kinetic energy of the system comes from large scale accretion processes. 
As mentioned above, after a shocked is produced, the turbulent environment of the gas dynamics facilitates the gas to 
reach high density-low temperature states. Under such conditions the gas is able to locally collapse 
gravitationally transforming their gravitational energy into kinetic energy at small scales producing 
the feature at high wave numbers in the power spectrum.

The distribution of energy in solenoidal and compressive modes shows that the solenoidal modes content 
$R_k\approx 2/3$ of the total kinetic energy. This energy distribution is the consequence of a random 
turbulent motion which for each compressional direction, e.g. $x$, presents two rotational directions, 
$y$ and $z$: the kinetic energy follows an equipartition rule with the same amount of energy for each 
direction. This energy distribution is not constant neither in time nor in scale but it has well stated 
values $65\la R_k\la70$ below $\sim$100 pc. 

All the Fourier information presented above can be very useful in order to set initial conditions for idealized primordial star formation simulations in the spirit of \citet{Clarketal2011}. For instance, these authors take an initial velocity power spectrum with a slope -4. This value is much lower than our slope $\approx-2$. The inclusion of a -2 slope in the previous study could bring a more efficient fragmentation at small scales with a possible more efficient low mass primordial star formation
% Referee 2º comments
as found in \citet{Clarketal2011b} and \citet{Greifetal2011} who, using 3D numerical simulations from cosmological initial conditions, have shown that a multiple system of low mass ($\sim 0.1-10$M$_{\odot}$) primordial stars can be formed inside DM mini halos.  

Using a one zone model \citet{Schneideretal2011} have shown that
without dust there is a critical metallicity ($Z\sim
10^{-4}Z_{\odot}$) above which gas can fragment into
$\ge10$M$_{\odot}$ object. Furthermore, they found that there is a
critical dust-to-gas ratio above the gas can fragment into low mass
($\sim 0.01-1$M$_{\odot}$) objects. The existence of a critical
metallicity above which it is possible to form low mass star is
explained by the low-mass star-formation critical metallicity theory
of \citet{BrommLoeb2003}. \citet{Frebeletal2007} tested this theory
with data of stars formed at the galactic halo, globular clusters and
dwarf spheroidal galaxies reaching a good agreement among data and
theory. Nevertheless, the solidness of this theory has been questioned
by a recently found extremely metal poor star which seems to violate
the theory \citep{Caffauetal2011} in the sense that it do not has the
amount of both oxygen and carbon needed to form a low-mass star as
stated in the \citet{BrommLoeb2003} theory. This finding makes
possible the potential observation of low mass primordial stars as
mentioned in \citet{JohnsonKhochfar2011}.

% Third minor point
The study presented in this work does not take into account the possible radiation background from previously formed pop III stars. Of course, a high enough number of photons in the Lyman-Werner band could avoid the gas fragmentation due to the molecular coolants destruction \citep{Haimanetal1997}. The result of this process could be non fragmented hot gas piled on the main halo center of mass. This configuration could be responsible for the formation of super massive black holes at high redshifts (e.g. \citep{Shangetal2010}).  
% Fourth minor point

Our work either take into account the possible supernovae explosion from pop III stars previously formed inside the merged mini halos. For masses of the super novae progenitor in the range $15-40$ M$_{\odot}$, this kind of violent phenomena certainly would be able to expel the gas inside the DM mini halos, and for massive enough super novae progenitors, in the range of $140-260$ M$_{\odot}$, the explosion could be able to disrupt the gas inside halos even more massive than $\sim10^7$ M$_{\odot}$ \citep{Whalenetal2008}. Because our main target is to try to understand the main processes involved in the formation of supersonic turbulence in unperturbed primordial environments we have neglected these two important feedbacks.

% Second minor point
Our cosmological initial conditions were set with $\sigma_{8}=0.9$. This value is larger than the most recent measure of $\sigma_8=0.8$ which imply that the DM halos in our simulation collapse faster than they do in a more realistic scenario. Actually, in a more realistic scenario the halo collapse process would occur at a redshift $z\approx 0.89 (1+Z_{0.9})-1$, where $Z_{0.9}$ is the collapse redshift with $\sigma_{8}=0.9$. For instance, the results that we show at z=10 should occur at $z\approx8.8$.  
\section{Summary and Conclusions}
We have performed hydrodynamical simulations of primordial gas from cosmological initial conditions in order to study how the DM merging process triggers the formation of turbulence in primordial environments and how the developed supersonic environment favors the formation of molecular coolants. 
The simulation, inside a 1Mpc/$h$ box size, follows the evolution of a $3\times10^7$ M$_{\odot}$ 
from $z\approx120$ to z=10 with a proper distance resolution $\Delta x\approx1.95$ pc at this redshift,
covering almost 5 orders of magnitude in distance scales.

Our main conclusions can be summarized as follows:

\begin{enumerate}
\item Both the baryonic accretion process through filaments and the DM minihalo merging onto the main 
halo is able to produce a supersonic ($\mathcal{M}\ga10$) turbulent and shocked ($v_{rms}\ga22$ km/s) 
environment where the H$_2$ and the HD molecules are formed efficiently.
\item The non equilibrium molecular formation triggered by the supersonic turbulent environment creates a number of regions with temperature $T\la300$ K. Some regions with number density $50\la n\la10^3$ cm$^{-3}$
reach temperatures below 100 K evidencing the effect of the HD molecule as the main coolant. At higher 
densities the H$_2$ is the more efficient coolant. All these low temperature-high density regions are 
potential star formation places. 
\item The Fourier analysis of the velocity field shows that the kinetic energy has a nearly Burgers 
spectrum $\propto k^{-2}$, albeit a bit steeper at lower redshift with a minimum exponent of $\approx$-2.2. 
This power law confirms the previous statement of supersonic and shocked environment. This spectrum 
approximately implies a $S_2(\ell)\propto\ell^{1.0}$ second order longitudinal structure 
function, steeper than the $S_2(\ell)\propto\ell^{0.76}$ behavior observed in local molecular clouds. 
\item The characterization of the turbulence power spectrum shows that it could favor the formation of low mass primordial stars.  
\item The energy spectrum does not decay at high wave numbers as fast as shown in isothermal simulations 
without self-gravity. This behavior is explained by taking into account that after the accretion process 
enhances the cooling at supersonic turbulent regions the gas is able to form gravitationally unstable over 
densities. In this way, the local gravitational collapse enhances the velocity power at scales of few 
$\times10$ pc converting the gravitational energy into kinetic energy. 
\item The energy in solenoidal (rotational) modes is $R_k\approx2/3$ of the total energy as in previous simulations of turbulent gas. 
\end{enumerate}

This paper is part of our ongoing effort \citep{jpp2} to produce numerical simulations that capture enough physics to understand the process of galaxy formation at the highest redshifts in order to shed light on what mechanisms could be dominating and shaping the future evolution of galaxies. We have shown here that the presence of relatively low mass halos ($\sim10^7$ M$_{\odot}$) dramatically influences its environment by creating a turbulent ISM. This in turn makes the smaller mass halos ($M \sim 10^6 M_{\odot}$), the most abundant ones at that redshift, to also develop turbulence in their ISM because of the merging process. The end result is that the production of coolants is enhanced, so much that even the HD molecule becomes an important coolant in some regions, thus producing regions of high density-low temperature that could be sites of star formation. We are now investigating how universal the process presented in this paper is by analyzing the whole 1Mpc box simulation and including the low mass and high mass dark matter halos. We will present our findings in a forthcoming paper. We can at this point speculate that, if this mechanism turns out to be universal at such high redshift, it could also potentially work at lower redshifts, so that the initial turbulence in the ISM of a galaxy could be due to the merging process, which is universal in the standard LCDM scenario.

\end{document}